\documentclass[preprintnumbers,amsmath,amssymb,twocolumn,floatfix,prl]{revtex4}

\usepackage{dcolumn}
\usepackage{bm}
\usepackage{graphicx}

\begin{document}     
\title {Incompressible Quantum Hall fluids as
  Electric Quadrupole fluids}
\author{ F. D. M. Haldane}
\affiliation{Department of Physics, Princeton University,
Princeton NJ 08544-0708}
\date{February 28, 2023, revised April 2, 2023}
\begin{abstract}
 A new picture of both integer and fractional incompressible
 quantum Hall fluids as fluids carrying an electric quadrupole is introduced.
 This clarifies their geometric properties, provides a new generic expression
 for Hall viscosity, and allows removal of the ingredient of continuous rotational
 symmetry (absent when electrons move in a crystalline background) from their description.
  \end{abstract}
\maketitle
Much of the theory of the quantum Hall (QH) effect, both integer (IQH) and
fractional (FQH), has been based on models with a continuous ($SO(2)$) rotation
symmetry  around the normal to the two-dimensional (2D) flat ``Hall
plane'' that supports the
``incompressible quantum fluid''\cite{laughlin} that exhibits the effect.
However,  electron QH fluids are  physically realized in structures
aligned with flat
crystalline lattice planes which host a microscopic-scale 2D Bloch
bandstructure supporting Landau quantization,
and generically have no such continuous symmetry, so the status
of $SO(2)$ symmetry can only be that of a ``toy model" feature that simplifies
calculations, but should not be part of  any fundamental
description.

Many results superficially obtained using  $SO(2)$ symmetry, such
as the seminal results of Girvin, MacDonald and Platzman\cite{gmp} (GMP)
turn out on examination to only require a weaker symmetry, 2D
inversion (equivalent to $180^{\circ}$ rotation) in the Hall plane, which can be
identified\cite{fdmhprep} as the fundamental point symmetry of uniform
QH fluids at rest.
Exceptions are results on ``Hall viscosity''\cite{asz,readvisc}
(and related work on an extension of the description of QH fluids to surfaces with Gaussian curvature\cite{wenzee}), in
which local continuous rotation symmetry appears to enter in a fundamental way,
so these results
cannot be regarded as part of a generic description.

Here an apparently-previously-unrecognized fundamental feature of QH
fluids, an intrinsic \textit{electric quadrupole density}, is shown
to both provide a
reinterpretation of  these previous results, and their extension to remove the
fictitious ingredient of $SO(2)$ symmetry.     There are a number of variant
definitions of  the ``electric quadrupole'' tensor of a charge distribution
$\{q_i, \bm x_i\}$
used in the literature\cite{nist}: the
version used here, and taken as the fundamental one, is the
``primitive'' form
\begin{equation}
  q^{ab} = \tfrac{1}{2}\sum_i q_i (x^a_i - \bar x^a)(x_i^b-\bar x^b)
  \label{qpole}
\end{equation}
where $\bar {\bm  x}$ is the center of charge, $\sum_iq_i\bm x_i$ =
$\bar {\bm x}\sum_i q_i$.   (This differs from the 
``traceless'' form used in multipole expansions of solutions
of a Laplace equation, where  ``trace''  means the
contraction $\delta_{ab}q^{ab}$ with the Euclidean metric $\delta_{ab}$
used to define the Laplacian.)
The use of the ``primitive''  as opposed to ``traceless''
electric quadrupole distinguishes this present work from previous
work\cite{maciejko}, which considered a ``quadrupole'' to be the  (``traceless'')   
\textit{``nematic'' deviation} $\delta q^{ab}$ from rotational invariance, where $q^{ab}$ =
$q_0\delta^{ab} + \delta q^{ab}$.
Here it is proposed that  ``flux attachment''   to form composite
particles  leads to an emergent  
\textit{primitive} ``guiding center'' electric quadrupole density
which is a fundamental feature of incompressible FQH states.

In a fixed crystalline background, the only role  for the Euclidean metric in
low-energy descriptions of condensed matter (other than in
defining a Cartesian ``laboratory frame'' coordinate system
$\bm x$ = $x^a\bm e_a$, $\bm e_a\cdot \bm e_b$ = $\delta_{ab}$) is in
the Newtonian inertial-mass tensors $m\delta_{ab}$  of the atomic
nuclei in phonon dynamics, which is not considered here,
and (in contrast to earlier work\cite{haldane, gromov})
a ``\textit{no}-metric'' description of QH fluids in Landau levels
on flat lattice planes is presented.
Implicit use of a spatial metric is exposed by distinguishing
upper (contravariant) indices of displacements $x^a$ and
lower (covariant) indices of derivatives $\partial_a$ $\equiv$
$\partial/\partial x^a$; only mixed-index tensors such as the stress
tensor $\sigma^a_b$ can have a trace $\sigma^a_a$ (defined with index
summation convention).  Without
rotational symmetry to make it symmetric, the Cauchy form of the stress tensor
``$\sigma_{ab}$'' = $\delta_{ac}\sigma^c_b$ loses its meaning.
Index placement distinguishes the Euclidean
metric, its inverse $\delta^{ab}$, and the Kronecker symbol
$\delta^a_b$, which only coincide in Cartesian
coordinates.

With the definition (\ref{qpole}),
the potential energy of a neutral  fixed electric quadrupole $q^{ab}$  in an electric
field $\bm E(\bm x)$ is
 $-q^{ab}\partial_aE_b(\bar {\bm x})$, analogous
to that of a fixed electric dipole $p^a$, $-p^aE_a(\bar{\bm x})$.   A fluid
of charged electric quadrupoles with a spatially-varying quadrupole
density  $Q^{ab}(\bm x)$ has an electric polarization density $P^a(\bm
x)$ = $\partial_bQ^{ab}$, and a ``bound charge'' density
$-\partial_a\partial_bQ^{ab}$ in addition to the ``free charge''
density.   At the edge  $\bm x(\bm s)$ of such a fluid with outward unit normal
$\hat{\bm n} (\bm s)$, there is a surface polarization $\bm P
\cdot \hat{\bm  n}$ = $Q^{ab}(\bm x(\bm s))\hat {n}_a(\bm s)\hat
n_b(\bm s)$: the identification of the QH electric quadrupole density
made here immediately leads to a complete account of both IQH and FQH
edge dipoles.  The in-plane electric quadrupole density of a ``clean
limit''
incompressible 2D QH fluid becomes important because the fluid's fundamental
ground-state ``incompressibility'' property is  unbroken 2D inversion
symmetry\cite{fdmhprep},
with a bulk energy gap for 
excitations carrying electric polarization tangent to the Hall plane.

A uniform magnetic flux density through the Hall
plane will be represented by the in-plane antisymmetric
Faraday tensor $F_{ab}$ = $\epsilon_{ab}B$, where the matched
Levi-Civita symbols $\epsilon_{ab}$ and $\epsilon^{ab}$ have 
orientation  (handedness)  usually chosen so that
$eB$ = $\hbar/\ell^2$  $>$ 0  ($e$ is the electron charge).
When there is Landau quantization of electron motion in
 the plane, forming quasi-degenerate Landau levels (LLs),
 the electron coordinates $\bm x_i$ can be decomposed as
  $\bm R_i$ + $\tilde {\bm R}_i$, where $\bm R_i$ is the time-averaged
  center, or
  \textit{guiding center} (GC), of the  Landau
  orbit (LO),
  and $\tilde {\bm R}_i$  is the LO displacement $\bm x_i-\bm R_i$.
  In the absence of an electric field in the plane, the guiding center
  is static, $\langle \tilde {\bm R}_i\rangle$ = 0, and the electric
  dipole moment of the LO relative to its GC vanishes.   Landau quantization
  occurs when the magnetic flux through a microscopic unit cell of
  the Hall surface is much smaller than the flux quantum $2\pi
  \hbar/e$, and  suppresses
 electronic Umklapp processes, allowing a continuum description
  where $\hbar$ times a Bloch vector becomes a true momentum, and
  LLs become extensively degenerate.
 In the IQH effect, the only contribution to the
 (extensive) total quadrupole moment  is the expectation value of the
 LO term :
 \begin{equation}
   q_{\text{LO}}^{ab} = e\tilde {\Lambda}^{ab}, \quad \tilde \Lambda^{ab} =
   \tfrac{1}{4}\sum_i
   \{\tilde R^a_i,\tilde R^b_i\}.
 \end{equation}

It is here argued that the spontaneous
formation of  a GC contribution to this quadrupole
density, that lowers the correlation energy, is the fundamental mechanism
in the creation of FQH states, and stabilizes ``flux
attachment''.   This is given by $q^{ab}_{\text{GC}}$ = $e\Lambda^{ab}$,
\begin{equation}
  \Lambda^{ab} = 
  -\tfrac{1}{4}\sum_i \{R^a_i,R^b_i\}
  + \tfrac{1}{2}\int d^2\bm x \rho_0(\bm x)x^ax^b,
  \label{gcq}
  \end{equation}
where $\rho_0(\bm x)$ is the uniform ground-state GC
density $\nu/2\pi\ell^2$ inside the fluid with LL  filling $\nu$, and
zero outside.
The second term in (\ref{gcq})  removes a super-extensive
background term leaving an extensive negative-definite tensor $\Lambda^{ab}$
($\tilde \Lambda^{ab}$ is positive-definite) (formally, removing the background replaces
$\bm R_i$ by $\delta \bm R_i$ = $\bm R_i - \langle \bm R_i\rangle$).

The GC and LO vectors obey conjugate Heisenberg
algebras $[\tilde R_i^a, \tilde R_j^b]$ = $[R_i^b,R_j^a]$= $i(\hbar
e/B)\epsilon^{ab}\delta_{ij}$, with $[R^a_i,\tilde R^b_j]$ = 0.
Both $\tilde \Lambda^{ab}$ and
$\Lambda^{ab}$ obey the same $SO(2,1)$ Lie Algebra, presented as
\begin{equation}
  [\Lambda^{ab},\Lambda^{cd}] = \tfrac{1}{2}i \tfrac{\hbar}{eB}
  \left (\epsilon^{ac}\Lambda^{bd} + \epsilon^{bd}\Lambda^{ac} +
    a\leftrightarrow b\right ).
  \label{lie}
\end{equation}
The generators  $\Gamma^a_b$ of uniform strain, obeying
$[x_i^a,\Gamma^b_c]$ = $i\hbar \delta^a_cx_i^b$ and
$[\Gamma^a_b,\Gamma^c_d]$ = $i\hbar (\delta^a_d\Gamma^c_b -\delta^c_b\Gamma^a_d)$, 
are given by
$eF_{bc}(\Lambda^{ac} + \tilde \Lambda^{ac})$ + $\tfrac{1}{2}D\delta^a_b$, where
$D$ $\equiv$ $\Gamma^a_a$ = $\sum_ieF_{ab}R^a_i\tilde R^b_i$ is the dilatation operator,
$[D,\Gamma^a_b]$ = 0; $F_{bc}\Lambda^{ac}$ and $F_{bc}\tilde
\Lambda^{ac}$ are traceless.    An Abelian $SO(2)$ symmetry would be generated
by azimuthal angular momentum $L$, defined so that
$[L, x^a]$ = $i\hbar \epsilon^{ac}\delta_{bc}x^b$ where $\delta_{ab}$
is the Euclidean metric, so $L$ = $-eB\delta_{ab}(
\Lambda^{ab} + \tilde \Lambda^{ab})$.

The incompressible QH fluids (unlike Euler's 
causality-violating classical incompressible fluid model)
have no autonomous low-energy degrees of freedom.
Their flow velocity in the Hall plane is the drift velocity
$\bm v_D$, determined by the local tangential electric field:
$E_a(\bm x) + F_{ab}v^b_D(\bm x)$ = 0.   While the Euler fluid 
has an infinite speed of sound and equilibrates pressure
throughout its bulk  instantaneously,
the pressure in incompressible QH fluids vanishes in the  $T$ $\rightarrow$ 0
limit, when the stress tensor
$\sigma^a_b(\bm x)$ in  a flowing fluid  becomes traceless.
In a fluid, viscous stress is the response to a non-uniform flow velocity,
which in the case of QH fluids is the response to a non-uniform
in-plane electric
field that couples to  local electric quadrupole density.

The generators $\Gamma^a_b$ of uniform strain can be used
to obtain the dissipationless Hall viscosity in the
ground state of an incompressible (gapped) quantum fluid using the
expression for the ``quantum Lorentz force''  $f_{\mu}$ = $-\langle
\Psi|\partial_{\mu}H|\Psi\rangle$ of adiabatic
quantum mechanics\cite{asz}:
\begin{equation}
  f_{\mu}(\bm {\mathrm x}) = -\partial_{\mu} E(\bm {\mathrm x}) +
  \hbar \mathcal F_{\mu\nu}(\bm {\mathrm
  x})\dot {\mathrm x}^{\nu} ,\quad \dot{\bm{\mathrm x}} \equiv \partial_t
\bm {\mathrm x},
  \label{lor}
\end{equation}
where $E(\bm {\mathrm x})$ is the non-degenerate  eigenvalue of an
eigenstate $|\Phi(\bm {\mathrm x})\rangle$ of a
Hamiltonian $H(\bm {\mathrm x})$, where $\bm {\mathrm x}$ = $\{\mathrm
x^{\mu}\}$ is a set of external \textit{control parameters}, with
$\partial_{\mu}$ $\equiv$ $\partial/\partial {\mathrm x}^{\mu}$,
which are sufficiently-slowly
varied along a path $\bm {\mathrm x}(t)$ in parameter space so that,
under the action of $H(t)$ = $H(\bm {\mathrm x}(t))$,
a state $|\Psi(t)\rangle$, given initially at $t$ = $t_0$ by  $|\Phi (\bm
{\mathrm x}(t_0))\rangle$, remains close to (has an infinitesimal  quantum
distance from) $|\Phi (\bm {\mathrm
  x}(t))\rangle$ at later times.
Here  $\mathcal F_{\mu\nu}$ is the Berry curvature $\partial_{\mu}\mathcal A_{\nu}
  -\partial_{\nu}\mathcal A_{\mu}$, with $\mathcal A_{\mu}$ = $-i\langle
  \Phi|\partial_{\mu}\Phi\rangle$.
  
The viscosity tensor $\eta^{ac}_{bd}$ is defined by the linear response of stress to
the velocity gradient (rate of strain) of a fluid:
 $\sigma^a_b(\bm x)$ = $\eta^{ac}_{bd} (\bm x)\partial_cv^d(\bm
 x)$. The expression (\ref{lor}),
where $\bm {\mathrm x}$ $\mapsto$ $\bm e$ parametrizes a strain,
was used in \cite{asz} to obtain an expression equivalent (after minor
corrections and reformulation  using $H(\bm e)$ = $U(\bm e) H U(-\bm e)$,
with $U(\bm e)$ = $\exp (i e^b_a\Gamma^a_b/\hbar$)) to
  \begin{equation}
    \int d^d\bm x \, \eta^{ac}_{bd}(\bm x)
    = -i\hbar^{-1}\langle \Psi|[\Gamma^a_b,\Gamma^c_d]|\Psi\rangle ,
    \label{visc}
  \end{equation}
which is $\hbar$ times the Berry curvature with respect to strain.
Ref. \cite{asz} proceeded to evaluate the Berry curvature assuming  Newtonian
mechanics, with the Euclidean metric in the mass tensor as the
parameter modified by strain, $\delta_{ab}$ $\mapsto$ $g_{ab}$, $\det
g$ = 1,
suggesting to some an interpretation (\textit{not} endorsed here)
 of Hall viscosity as a
``response to variation of background geometry".

Since the trace $\Gamma^a_a$ is proportional to the dilatation operator which
commutes with $\Gamma^c_d$, (\ref{visc})  is
traceless with respect to both the left pair and right pair of
indices, and odd under their interchange (thus dissipationless).
Using the Lie algebra of $\Gamma^a_b$, the Hall viscosity of a QH fluid is
\begin{equation}
  \eta^{ac}_{bd}(\bm x) = \delta^a_dF_{be}Q^{ce}(\bm x) -
  \delta^c_bF_{de}Q^{ae} (\bm x),
  \label{newformula}
\end{equation}    
where $Q^{ab}(\bm x)$ is the local electric quadrupole density, which in 2D has
dimensions [electric charge].

 Both IQH and FQH fluids have a natural interpretation in terms of 
an elementary unit, the ``composite boson"
(CB)\cite{girvinbook, girvinmac87,zhk,read89} that couples both to an
Abelian $U(1)_k$ ``(2+1)-D" emergent Chern-Simons (CS) gauge field and the
electromagnetic field. In the IQH case, the CB is the composite of an
empty orbital in an otherwise-filled LL with an electron
that is added to occupy it.   Interchange of two such entities in
the Slater-determinant state of the filled LL is a
row interchange followed by a column interchange, so is bosonic (no
phase change).

The difference between orbitals in a LL
and the Wannier-like orbitals in a usual band insulator is that
former can be adiabatically moved, while the latter are bound to the
lattice.  The Berry phase generated as the GC of the empty
orbital moves
through the filled background
is described as an emergent ($|k|=1$) CS gauge field 
that cancels the Aharanov-Bohm phase of the electron it contains when
filled, making the CB  effectively neutral and able to condense.

Whereas the CB description is merely formally interesting
in the IQH case, it becomes central in the FQH case, when the CB
(which has strong analogies to the unit cell of a solid) is
formed from a set of $p$ particles added to a group of $q>1$ empty
orbitals, that can be thought of as an empty ``bubble'' or void in the
fluid background, forming the elementary unit of the fluid.
The ``bubble'' represents quantized ``attached flux'', reinterpreted as
attached one-electron orbitals: short range repulsion between
the particles bound inside it, and those outside, favors placing
the bound particles close to the center, creating a quadrupolar charge
distribution of guiding centers inside it with total CB charge $e^*$ =
$pe$.
The shape of the void
and its contents will adjust to find the quadrupole giving the
lowest correlation energy of the fluid.

The Berry
phase as the void moves is described by a  Abelian CS
gauge field with a signed integer index $k$ where
$| k|$ = $pq$.
Exchange of two voids produces  a (conventional)
exchange factor $(-1)^k$ (this requirement quantizes $k$ to integer values
in Abelian CS theory).  For the composite to be a boson, this factor
must cancel the exchange factor of the group of $p$ particles inside it
(\textit{i.e.}, $(-1)^k$ = $(-1)^p$ if the particles are electrons).

The phenomenology of the QH fluids can be entirely described by the
composite-boson charge $e^*$ = $pe$, the level $k$ of its CS field, and
its electric quadrupole density $Q^{ab}$, without reference to the
microscopic description in terms of $p$ and $q$:
the Hall
conductivity $\sigma_H^{ab}$ is $e^{*2}\epsilon^{ab}/2\pi \hbar k$
(the sign of $k$ is orientation-dependent, that of $k\epsilon^{ab}$ is not),
and the elementary fractional charge is $\pm e^*/k$.
The composite boson also has a parity $\xi$ = $\pm 1$ under 2D
inversion.

If there is rotational invariance around the normal to the
Hall plane, $\xi$ is given\cite{fdmhprep} by  $(-1)^{S + \frac{1}{2}k}$,  where
$\hbar S$ is the ``intrinsic orbital azimuthal angular momentum of the
composite boson'': $S$  is integer for even $k$, half-odd-integer for odd
$k$, and the electric quadrupole density is given in terms
of $S$ by
\begin{equation}
  Q^{ab} =  (e^* S/4\pi k) \delta^{ab}
\label{rotq}
\end{equation}
where $\delta^{ab}$ is the inverse Euclidean metric.
Without subtraction of the background term in (\ref{gcq}),
the angular momentum about its center of a circular fluid droplet
of $\bar N$ CB's with an unexcited edge
is  $ (\frac{1}{2}k\bar N^2 + S \bar N)\hbar$:
while
the signs of both $S$ and $k$ depend on orientation, their
ratio is orientation-independent.
Such a droplet can be placed on a sphere, centered at the north pole,
and the edge contracted to vanish at the south pole, covering the sphere
uniformly\cite{fdmhsphere}.
As $2S$ is integral, the internal angular momentum $\hbar S$
can be viewed as a 3D spin aligned along the outward normal of the
sphere at the position of the composite boson, which picks up the
standard spin-$S$ Berry
phase as the normal traces out a path on the unit sphere 
while the composite boson  moves on the surface.  This is an extrinsic
description equivalent to the intrinsic ``spin-connection'' picture of
Wen and Zee\cite{wenzee}.  The quantity $S$ = $p\bar s$ =
$-\tfrac{1}{2}p\mathcal S$ is related to previously-defined
quantities $\bar s$,``intrinsic orbital angular
momentum per electron''\cite{readvisc}, and $\mathcal S$,
``shift on the sphere''\cite{wenzee}: these are rationals, but \textit{not}
generically integers or half-integers.

A striking result\cite{wenzee} from the model of a QH fluid on a curved surface
is that there is an excess  charge density proportional to   $S$
times the Gaussian curvature $\kappa$.
Consider a simple example, the IQH in a filled  Newtonian Landau level with
index $n$:  the excess charge density on the sphere\cite{fdmhsphere} is
$e(n+\tfrac{1}{2})\kappa/2\pi$, and
the electric quadrupole density on the flat plane is easily evaluated
as $Q^{ab}$ = $e(n+\tfrac{1}{2})\delta^{ab}/4\pi$; both expressions
have no orientation ambiguity.
On a surface with
Gaussian curvature, a curvilinear coordinate system must be used,
but the metric $g_{ab}(\bm x)$ can still be
chosen unimodular, $\det g(\bm x)$ = 1.    The coordinate system can
be
chosen to be ``inertial''
at a chosen point $\bm x_0$, so  $g_{ab}(\bm x_0)$ =
$\delta_{ab}$, and the
first derivatives (Christoffel symbols) vanish: the Gaussian
curvature at $\bm x_0$ is then 
$-\tfrac{1}{2}\partial_a\partial_bg^{ab}(\bm x_0)$, 
consistent with the result that a spatially-varying electric
quadrupole density gives rise to a  ``bound charge''
density $-\partial_a\partial_bQ^{ab}$.

However, while the spherical geometry with rotational symmetry is a
useful toy model, there seems no physical basis for introduction
of Gaussian curvature: 2D Landau quantization occurs on lattice
planes with an underlying Bloch bandstructure: lattice planes in 3D
crystals are macroscopically flat, and graphene sheets, while
allowing extrinsic curvature, are (like a sheet of paper) profoundly resistant to Gaussian
curvature: if a graphene sheet is forced to conform to a curved
surface, the required Gaussian curvature will be localized at cone
singularities where folds intersect. 

A quadrupole density $Q^{ab}(\bm x)$ is here a symmetric definite tensor which
can be written as $Q_0(\bm x)g^{ab}(\bm x)$ where the field $g^{ab}$
is the analog of the inverse of a Euclidean metric with unit
determinant and positive signature.   It obeys a local analog of the
algebra (\ref{lie}) (which conserves $\det\Lambda$ = $\tfrac{1}{2}\epsilon_{ac}\epsilon_{bd}\Lambda^{ab}\Lambda^{cd}$, the  quadratic
Casimir), and its fluctuations are the fluctuations of $g^{ab}(\bm x)$
which conserve $Q_0(\bm x)$.    The field is analogous to the order-parameter
of a Heisenberg ferromagnet, except that, instead of  being mapped to the unit
sphere of the $SO(3)$ group, the quadrupole field is defined on one of
two  hyperboloids of the $SO(2,1)$ group\cite{so21}, mappings to the
hyperbolic plane, corresponding to  two possible signatures of $Q^{ab}$.

The
analog of a Zeeman term that energetically fixes the preferred
orientation of a ferromagnetic order-parameter is $-g^0_{ab}g^{ab}(\bm
x)$,
which is minimized when $g^{ab}(\bm x)$ is the inverse of $g^0_{ab}$.
However, $g^{ab}(\bm x)$ is physically unambiguous, and \textit{not} used as a
metric to set up a
coordinate system.  Since physical QH systems are defined on a flat lattice
plane, the laboratory-frame Cartesian coordinate system is used, with
no coordinate-reparametrization ambiguities, and no ``spin-connection'',
casting doubt on whether the  conjectured Wen-Zee term\cite{wenzee}
(and ``second Wen-Zee term''\cite{bradlynread,gromovfradkin}) 
really have a place in the (bulk) theory of the FQH unless they appear
after integrating-out the
gapped fluctuations of $g^{ab}(\bm x)$, \textit{i.e.},
long-wavelength fluctuations of the GMP collective mode\cite{gmp}.  This might occur
in  rotationally-invariant systems, where $g^0_{ab}$ is the
Euclidean metric $\delta_{ab}$, and the quadrupolar dynamics
are analogous to those of a ferromagnet with axial symmetry around the
direction of the Zeeman field, with a local conservation law for
$\delta_{ab}Q^{ab}(\bm x)$,
but in general, this will not be  conserved.

For electron-type FQH fluids, the
GC quadrupole has opposite signature to the 
LO quadrupole, but in hole-type fluids, such as the
particle-hole conjugate of the Laughlin state, they have the same
signature.   In a multiple-component Abelian FQH state\cite{fdmhsphere}, the fluid is
constructed hierarchically from $p$ simple sub-fluids, and each one can be
expected to have a dynamically-independent quadrupole field, some with
electron-type and some with hole-type signature.   By
analogy with the gapped magnons
of Zeeman-aligned ferromagnets with $p$ spins in the unit cell, one
can expect $p$ distinct GMP-like modes at long wavelengths.  The
sub-fluids may have different signatures (as in the $\nu$ = 2/7
state\cite{nguyen, boyang, papic}), somewhat  analogous
to ferrimagnets, where some sublattices have spin
 antiparallel to the dominant spin.

With rotational invariance, the quadrupolar
gapped ``magnon-like'' excitations of the FQH sub-fluids carry angular
momentum $\pm 2\hbar$, where those of electron-type sub-fluids
have the same sign (and double the magnitude) of that of the
inter-Landau-level
magnetoplasmon, while those of hole-type fluids have opposite sign.
(In the $SO(2,1)$ case,  ``magnons'' increase the quadrupole magnitude, in
contrast to the $SO(3)$ case, where magnons decrease the local spin magnitude.)
Without rotational invariance, angular momentum is not
conserved,  but these signs persist as chiralities.
A fuller discussion of the quadrupolar dynamics
will be given elsewhere.

In the absence of LL mixing, the GC
quadrupole is odd under a particle-hole  (p-h) transformation of the Landau
level that hosts the FQH fluid, so must vanish if p-h
symmetry is unbroken.
If the suggestion made here (that
the formation of the GC quadrupole by ``flux attachment''
is the key ingredient that drives the formation of FQH states) is
correct, unbroken p-h symmetry is
incompatible with the FQH effect, and a half-filled LL (HFLL) with
unbroken p-h symmetry and translational symmetry would have to be in a
gapless composite-fermion (CF)\cite{jain}
Fermi-liquid state (CFL)\cite{hlr,son}).      This is
consistent with a recent study\cite{sun} of the  HFLL in
Bernal-stacked graphene
bilayers, which show two different FQH
states with the signature (in the plateau spectrum near half-filling)
of broken p-h symmetry, or a gapless state (these states are
consistent with ``Pfaffian''\cite{mr}, ``anti-Pfaffian''\cite{apf1,apf2} and
CFL states.
If the conjectured principle is correct, there may be some analog
of the Lieb-Schultz-Mattis spin-chain theorem\cite{lsm} that requires gaplessness of
the HFLL with a translationally-invariant projected two-body
interaction between GCs if
p-h symmetry is not spontaneously broken (and translational invariance
plus 2D inversion is preserved).

While the gapped HFLL in graphene  bilayers is clearly seen to have  broken p-h
symmetry,  thermal Hall studies\cite{heiblum} of the $\nu$ = $5/2$ state in
GaAs-based heterostructures (which, after projection into the LL, have
a different two-body interaction between GCs)
appear to show a chiral Virasoro anomaly $c$ =
$\nu$ = 5/2, consistent with p-h symmetry of the partially-filled second
LL (the chiral Virasoro anomaly $c$ can be decomposed as $\nu + \tilde
c$, where $\tilde c$ can be obtained\cite{yeje} from the GC entanglement
spectrum\cite{li}, and vanishes when there is p-h symmetry). However, exact-diagonalization searches for a
particle-hole symmetric  ``PH-Pfaffian'' FQH state\cite{son} in a
single HFLL have to date proved fruitless (see also Ref.\cite{mross}).
LL mixing breaks p-h symmetry and  the degeneracy between Pfaffian and
anti-Pfaffian states: if the principle conjectured here is correct, a
FQH state with half-integral $c$ = $\nu$ requires sufficient
Landau-level mixing to make it energetically-preferred to both Pfaffian
and anti-Pfaffian (consistent with the findings of Ref.\cite{milovanovic}).

While there is a preferred $Q^{ab}$ which minimizes the correlation
energy in the uniform fluid, there will be a positive-definite deformation energy density $\delta u$ =
$\frac{1}{2}\chi^{-1}_{abcd}\delta Q^{ab}\delta Q^{cd}$, so the
response to a non-uniform electric field is given by $\delta Q^{ab}$ =
$\chi^{abcd}\partial_cE_d$, leading to a bound-charge density
$-\partial_a\partial_b\delta Q^{ab}$ that will
spread the point charges of topological excitations
into their immediate environment, analogous to the effects of skyrmions in quantum
Hall ferromagnets\cite{sondhi}, or modify the profile of the fluid at edges, where
strong field gradients and flow-velocity gradients occur.

In summary,  an emergent (primitive) guiding-center electric
quadrupole density
is identified as a
key universal property of FQH liquids, adding to a Landau-orbit electric
quadrupole density of IQH liquids, to  provide both a new fundamental
formula (\ref{newformula}) for Hall viscosity, and new ways of
understanding  their energetics and gapped collective excitations,
without invoking a commonly-imposed  but unphysical continuous $SO(2)$ 2D rotational invariance,
inappropriate in a fixed crystalline condensed matter background.

\begin{acknowledgments}
This research was primarily supported by NSF through the Princeton
University (PCCM) Materials Research Science and Engineering Center DMR-2011750.
\end{acknowledgments}


\begin{thebibliography}{2}
\bibitem{so21}
The connection of quadrupole fields to a $SO(2,1)$  analog of a
ferromagnet  was previously recognized in Ref.\cite{maciejko}, albeit in the
``traceless'' form  $\delta q^{ab}$  as a ``nematic fluctuation'' away
from $SO(2)$ rotational invariance.
\end{thebibliography}
\end{document}